\documentclass[reprint,superscriptaddress, nofootinbib,longbibliography]{revtex4-2}
\pdfoutput=1
\usepackage{graphicx}
\usepackage{amssymb}
\usepackage[linktoc=none]{hyperref}
\usepackage[dvipsnames]{xcolor}
\usepackage{dcolumn}
\usepackage[english]{babel}
\usepackage[utf8x]{inputenc}
\usepackage[T1]{fontenc}
\usepackage{eqnarray,amsmath,physics,bbold}
\usepackage{ dsfont }
\usepackage{appendix}
\usepackage{xcolor}
\usepackage{breqn}
\usepackage{comment}
\usepackage{physics}
\usepackage{relsize}
\usepackage{amsmath}
\usepackage{caption}
\usepackage[labelformat=simple]{subcaption}

\DeclareCaptionLabelFormat{subcaptionlabel}{\normalfont(\textbf{#2}\normalfont)}
\usepackage{hyperref}
    \hypersetup{
    colorlinks   = true,    
    urlcolor     = blue,    
    linkcolor    = blue,    
    citecolor    = red      
    }

\DeclareMathAlphabet{\mathpzc}{OT1}{pzc}{m}{it}

\newcommand{\INFN}{INFN, Sezione di Napoli, Gruppo Collegato di Salerno, I-84126 Napoli, Italy}

\newcommand{\UNISA}{Physics Department ``E.R. Caianiello'', University of Salerno, Via Giovanni Paolo II, 132, I-84084 Fisciano, SA, Italy}

\newcommand{\CNR}{CNR-SPIN, Via Giovanni Paolo II, 132,  I-84084 Salerno, Italy }

\newcommand{\IFW}{Institute for Theoretical Solid State Physics, IFW Dresden, Helmholtzstr. 20, 01069 Dresden, Germany}

\begin{document}
\title{Edelstein Effect in Isotropic and Anisotropic Rashba Models
}

\author{Irene Gaiardoni}
\email{igaiardoni@unisa.it}
\affiliation{\UNISA}

\author{Mattia Trama}
\affiliation{\UNISA}
\affiliation{\IFW}

\author{Alfonso Maiellaro}
\affiliation{\UNISA}
\affiliation{\CNR}

\author{Claudio Guarcello}
\affiliation{\UNISA}
\affiliation{\INFN}

\author{Francesco Romeo}
\affiliation{\UNISA}
\affiliation{\INFN}

\author{Roberta Citro}
\email{rocitro@unisa.it}
\affiliation{\UNISA}
\affiliation{\CNR}
\affiliation{\INFN}

\begin{abstract}
We investigate spin-to-charge conversion via the Edelstein effect in a 2D Rashba electron gas using the semiclassical Boltzmann approach. We analyze the magnetization arising from the direct Edelstein effect,
taking into account an anisotropic Rashba model. We study how this effect depends on  the effective masses and Rashba spin--orbit coupling parameters, extracting analytical expressions for the high electronic density regime. {Indeed, it is possible to manipulate the anisotropy introduced into the system through these parameters to achieve a boost in the Edelstein response compared to the isotropic Rashba model.}
We also discuss the theoretical framework to study the inverse Edelstein effect and calculate self-consistently the electric current induced by the proximity of the system to a ferromagnet. 
These results provide insights into the role of Rashba spin--orbit coupling and anisotropic effects in spin--charge conversion phenomena.
\end{abstract}

\maketitle

\begin{figure} [b!!]
\centering
\includegraphics[width=0.45\textwidth]{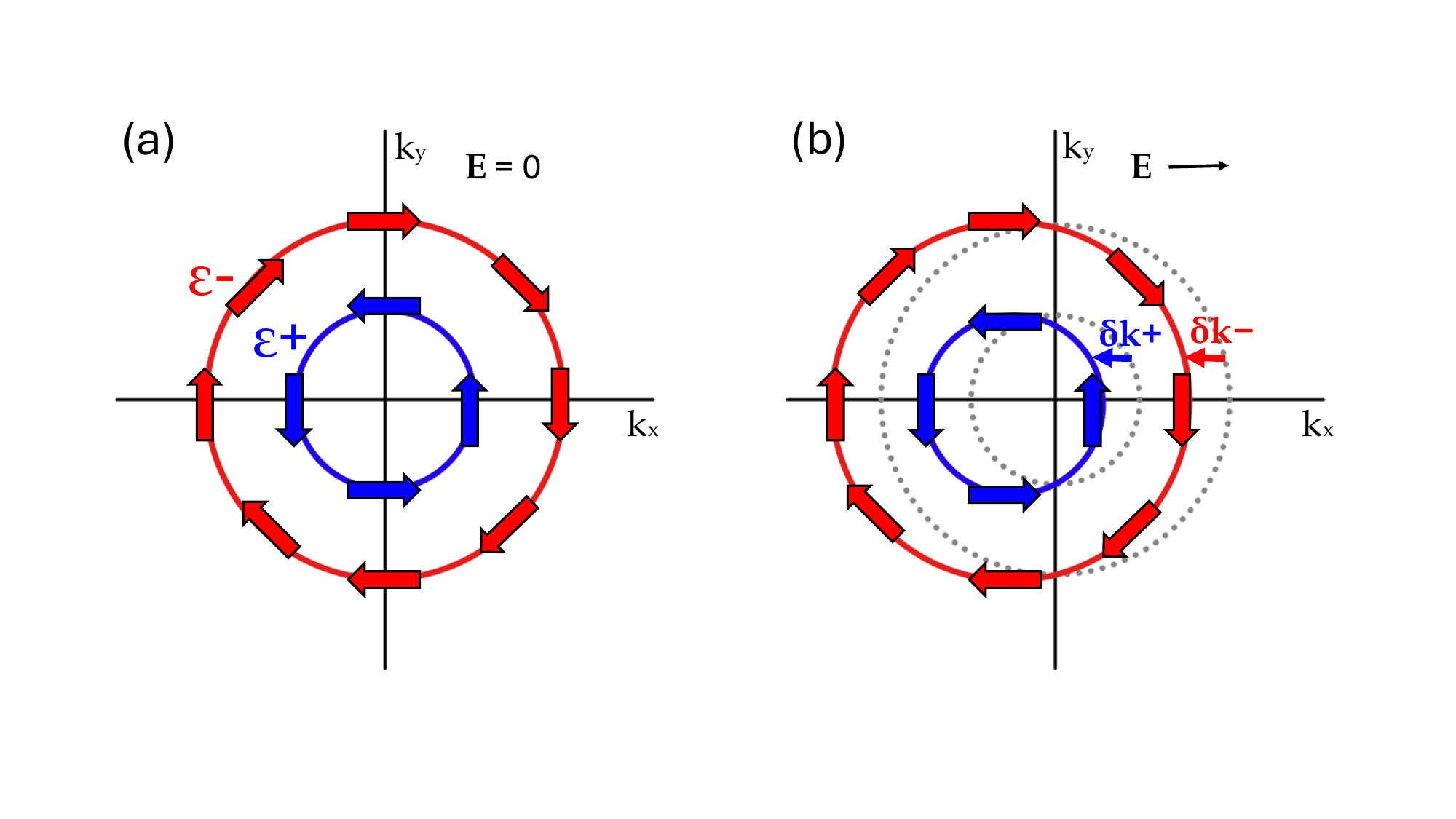}
 \caption{\raggedright Direct Rashba--Edelstein Effect: Blue and red indicate the branch of the energy dispersion, + or $-$. (\textbf{a}) In equilibrium, the total spin polarization vanishes. (\textbf{b}) If an external electric field$\mathbf{E}= E_x \hat{x}$ is applied, the Fermi lines are shifted opposite to the field direction, and anonvanishing spin polarization perpendicular to $\mathbf{E}$ results.}   
 \label{Edelstein}
\end{figure}

\begin{figure}[b!!]
\begin{subfigure}[t]{0.49\textwidth}
 \centering
 \hspace{-6pt}\includegraphics[width=\textwidth]{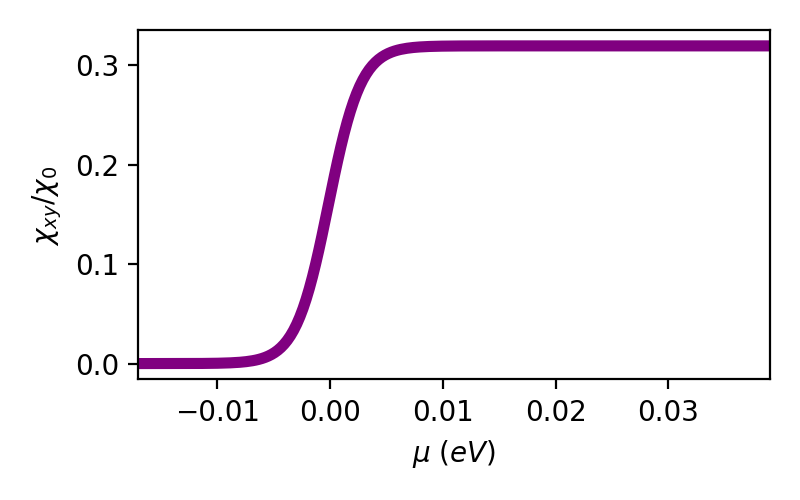}
 \label{0pannello-a}
 \end{subfigure}
 \begin{subfigure}[t]{0.40\textwidth}
 \centering
 \hspace{-6pt}\includegraphics[width=\textwidth]{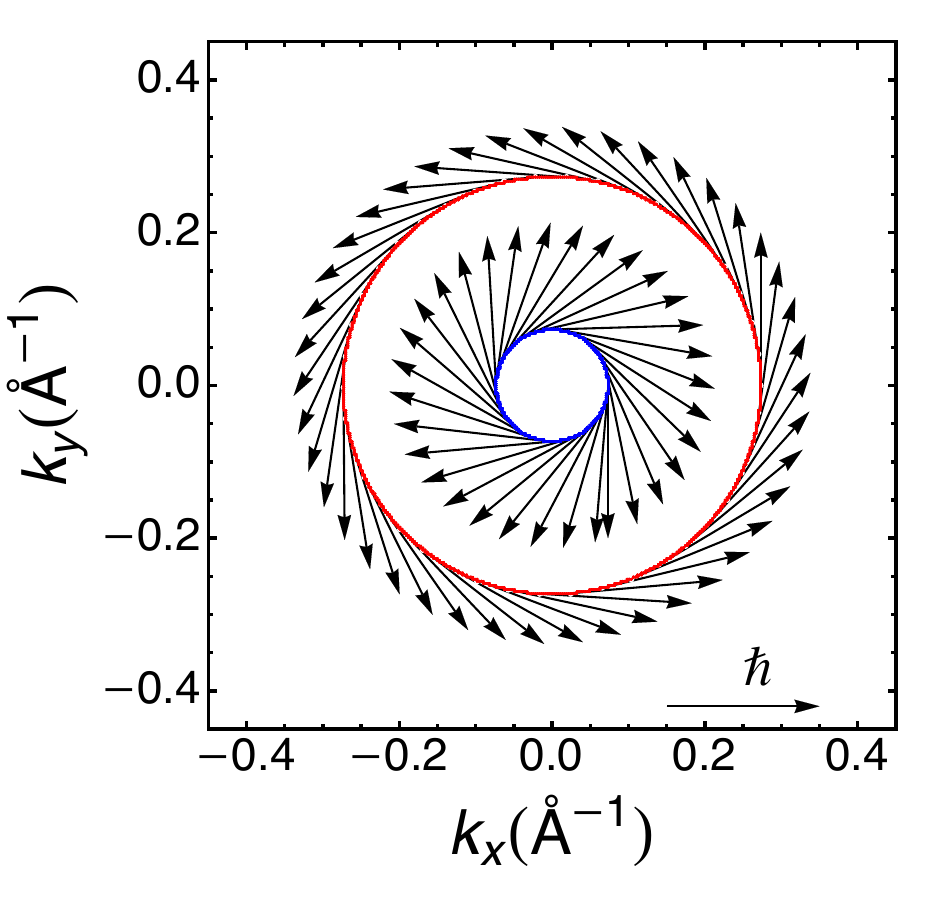}
 \label{0pannello-b}
 \end{subfigure}
 \caption{\raggedright \textbf{Left panel}: The figure shows the Edelstein susceptibility $\chi_{xy}/\chi_0$ ($\chi_0 = 2.44 \times 10^2 \, \mu_B \, \text{Å} \, \text{V}^{-1}$) as a function of the chemical potential $\mu$ for $\alpha = 52 \, \text{meV} \, \text{Å}$. \textbf{Right panel}: Fermi surface and spin structure for fixed chemical potential $0.06$~eV in an isotropic Rashba model. $\nu = +1$ indicates the inner circle (the blue one) and $\nu = -1$ the outer one (the red one).}
 \label{rashba_iso_2}   
\end{figure}

\begin{figure}[b!!]
    \begin{subfigure}[t]{0.49\textwidth}
    \centering \hspace{-6pt} 
       \includegraphics[width=\textwidth]{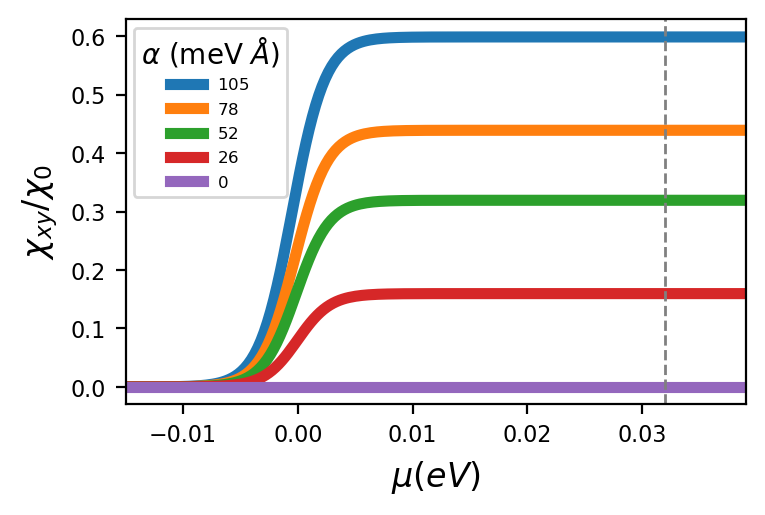}
        \label{pannello-a}
    \end{subfigure}
    \begin{subfigure}[t]{0.49\textwidth}
    \centering
        \hspace{-6pt} \includegraphics[width=\textwidth]{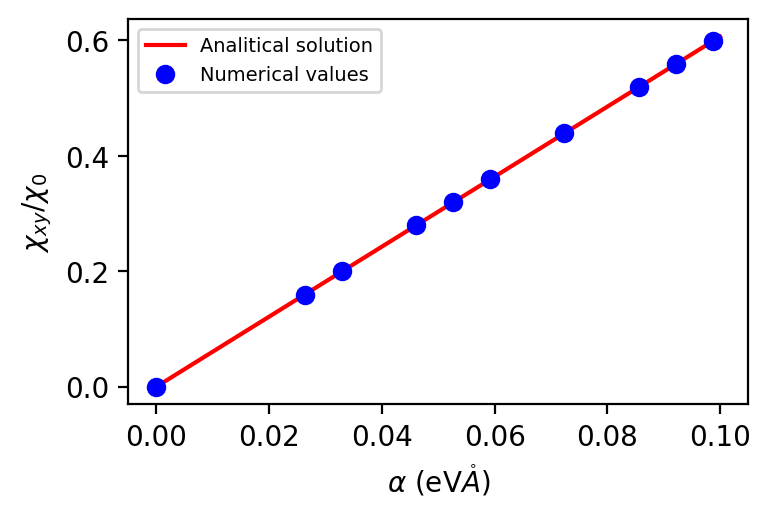}
        \label{pannello-b}
    \end{subfigure}
    \caption{ \textbf{Left panel}: Edelstein susceptibility $\chi_{xy}/\chi_0$ ($\chi_0= 2.44 \times10^{2}$ $\mu_B $ ~\AA~V$^{-1}$) as a function of the chemical potential $\mu$ for different values of $\alpha$. The dashed gray line indicates the fixed value of $\mu$ used for the plot in the right panel. \textbf{Right panel}: $\chi_{xy/} \chi_0$ as a function of $\alpha$ at fixed chemical potential $\mu= 3,32 \cdot 10^{-2}$ eV.}
    \label{rashba_isotropico 3}
\end{figure}

\begin{figure}[b!!]
    \begin{subfigure}[t]{0.43\textwidth}
    \centering
        \includegraphics[width=\textwidth]{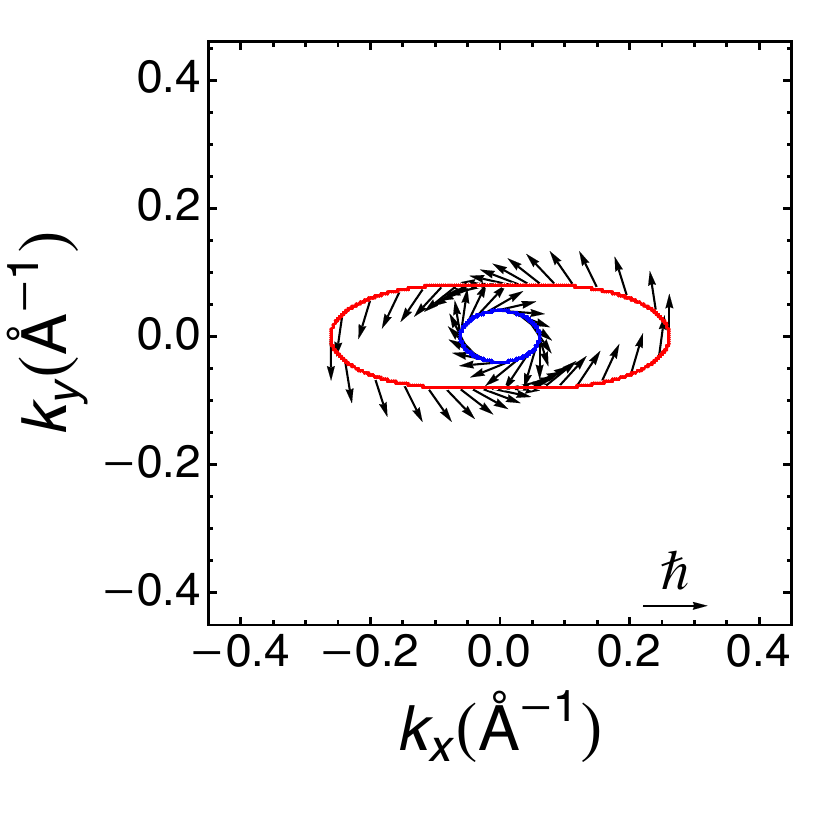}
        \label{pannello-ani-masse}
    \end{subfigure}
    \begin{subfigure}[t]{0.43\textwidth}
    \centering
        \includegraphics[width=\textwidth]{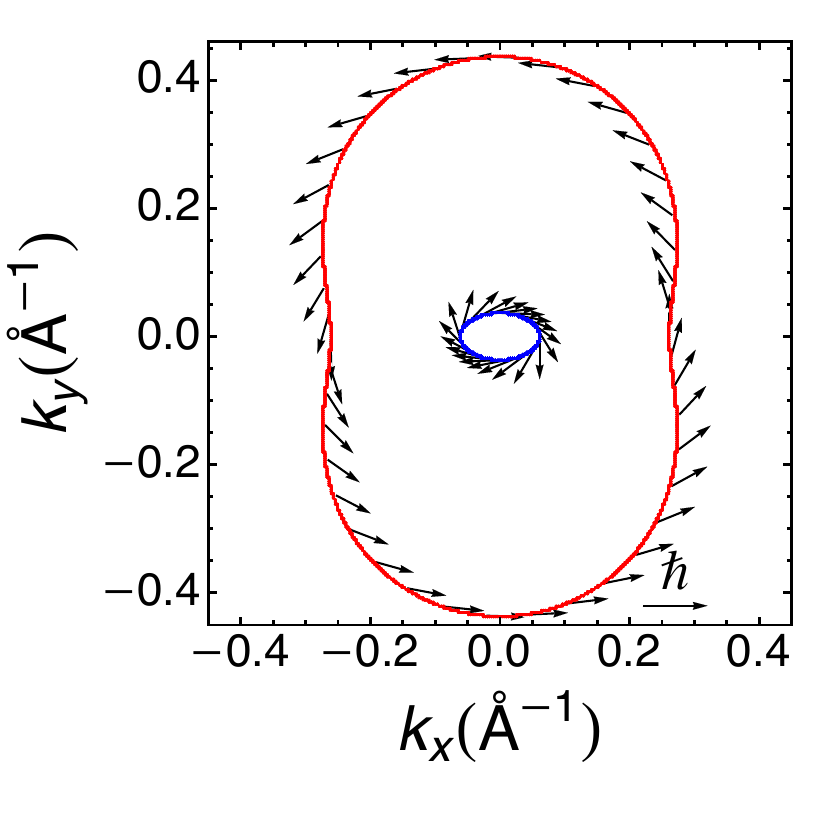}
        \label{pannello-ani-alpha}
    \end{subfigure}\vspace*{-6pt}
\caption{Fermi surface and spin texture for fixed chemical potential in the anisotropic case. \textbf{Left Panel}:   $m_y= 0.2 m_x$ and $\alpha_y=\alpha_x$. \textbf{Right Panel}: $m_y=m_x$ and $\alpha_y=2 \alpha_x$.}
\label{pannelli_ani}
\end{figure}

\begin{figure}[b!!]
    \begin{subfigure}[t]{0.49\textwidth}
        \includegraphics[width=\textwidth]{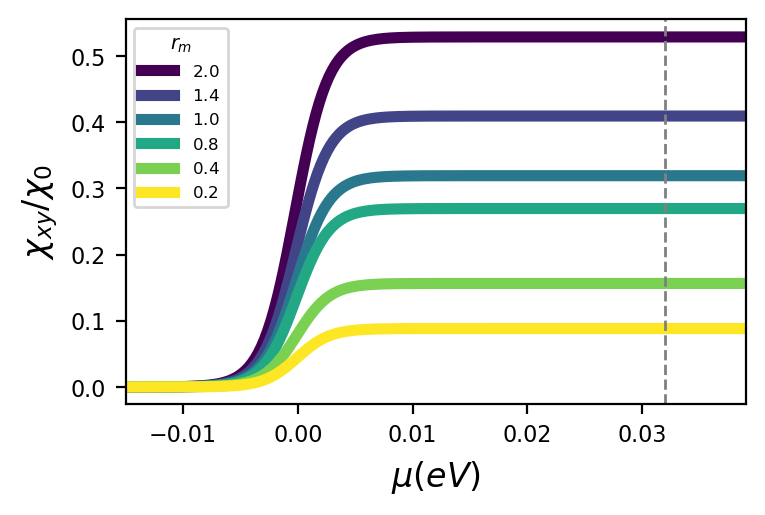}
        \label{2pannello-a}
    \end{subfigure}
    \begin{subfigure}[t]{0.49\textwidth}
        \includegraphics[width=\textwidth]{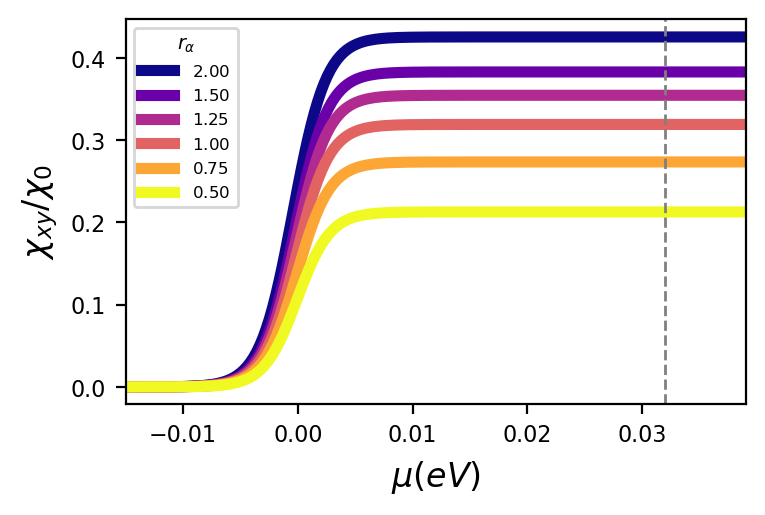}
        \label{2pannello-b}
 \end{subfigure}\vspace*{-6pt}
\caption{Edelstein susceptibility $\chi_{xy}/\chi_0$ ($\chi_0= 2.44 \times10^{2}$ $\mu_B$ ~\AA~V$^{-1}$) as a function of the chemical potential $\mu$ for different values of (\textbf{Left panel}) $r_m$, at $ \alpha_x=\alpha_y= 52$ meV \AA, and (\textbf{Right panel}) $r_{\alpha}$, at  $m_x=m_y$.}
\label{rashba_anitropico}
\end{figure}

\textit{Introduction}- Spintronics exploits the electron's spin degree of freedom, together with its charge, to develop advanced electronic devices, such as magnetic memories~\cite{wolf2001spintronics,schmidt2005concepts,soumyanarayanan2016emergent}. However, manipulating and detecting spin states often requires multiple external components, increasing the system complexity and the limiting efficiency~\cite{dieny2020opportunities}.
Spin--orbit coupling (SOC), which links the spin and orbital motion of electrons, offers an alternative approach by enabling spin control without the need for magnetic fields~\cite{go2021orbitronics,manchon2015new}. This principle underlies the emerging field of spin-orbitronics, which explores phenomena such as spin--momentum locking, spin torque, and topologically non-trivial spin textures, with the goal of realizing more compact and efficient devices.
Among the materials of interest, oxide interfaces, such as LaAlO$_3$/SrTiO$_3$ and LaAlO$_3$/KTaO$_3$, have attracted significant attention due to their inherently strong SOC and the ease of designing nanostructures. At these interfaces, a high-mobility two-dimensional electron gas (2DEG) forms due to the confinement potential~\cite{trama2023effect} acting on conduction electrons. These 2DEGs exhibit a variety of intriguing SOC-related phenomena, such as non-trivial Berry curvature effects~\cite{zhai2023large,trama2022gate,chen2025dirac} or topological superconductivity~\cite{guarcello2024probing,maiellaro2023hallmarks}, primarily driven by spin--momentum locking. This effect arises from the interplay between atomic SOC and broken inversion symmetry in the 2D system. Since the latter is induced by an electric field, these systems can be understood by the Rashba spin--orbit coupling (RSOC)~\cite{bychkov1984properties,rashba1960spin,10.21468/SciPostPhys.17.4.101}, whose Hamiltonian is
\begin{equation}
\hat{H}=\frac{p^2}{2 m}+\alpha \hat{z}\cdot(\textbf{p} \times \vec{\sigma}).
\label{hamiltonianaiso}
\end{equation}
The first term represents the kinetic energy, where $\mathbf{p}$ is the momentum and $m$ is the effective carrier mass. The second term corresponds to the RSOC itself, where $\alpha$ is the coupling strength, and $\vec{\sigma}$ is the vector of Pauli matrices. 
Defining the helicity operator as $\hat{S}=\hat{z}\cdot(\mathbf{p} \times \vec{\sigma})/ p$, with eigenvalues $s = \pm 1$, the spin--momentum coupling term induces an energy band splitting, with each band associated with a different helicity state. The resulting antisymmetric spin polarization in the electronic band structure, where spin remains tangential to the Fermi surfaces, as shown in Figure~\ref{Edelstein}a, makes these systems highly suitable for charge-to-spin conversion~\cite{amin2016spin, gariglio2018spin, vaz2019mapping, inoue2003diffuse,soumyanarayanan2016emergent, trier2022oxide, vicente2021spin, trama2022tunable, varotto2022direct, sinova2004universal, bibes2011ultrathin, gambardella2011current, trama2024non, seibold2017theory}.
\noindent In particular, such systems are expected to exhibit the Direct Edelstein Effect(DEE) ~\cite{edelstein1990spin,leiva2023spin, el2023observation,aronov1989nuclear,johansson2021spin,johansson2016theoretical}, also known as the inverse galvanic effect~\cite{gambardella2011current}. This phenomenon refers to the generation of an in-plane magnetization under an external electric field. Due to spin--momentum locking, a shift of the Fermi surfaces along $\hat{x}$ by $\delta k$ results in a spin imbalance in the Brillouin zone along the orthogonal $\hat{y}$ direction as shown in Figure~\ref{Edelstein}b.
However, for realistic materials, the intrinsic crystalline anisotropy, orbital degrees of freedom, and crystal distortions can drastically distort the Fermi surface even locally so that the simplified isotropic Rashba model is bound to fail. Thus, optimizing spin-orbitronics devices requires to go beyond this model, especially including the effects of anisotropies on the spin structure and the DEE.
In addition, the readout of the Edelstein signal is usually performed via the Inverse Edelstein effect (IEE)~\cite{shen2014microscopic,kato2004current, silov2004current, gorini2012onsager, sanchez2013spin, vignale2016theory}, the reciprocal counterpart of the DEE. This effect produces a charge current in response to an applied spin current. Despite their reciprocal nature, the DEE and IEE require different theoretical approaches, due to their distinct \mbox{initial conditions}.
Using a Boltzmann semiclassical approach, this paper provides an overview of the theoretical framework for the DEE and IEE in Rashba models. We first examine the dependence of the DEE on electronic density and investigate how the behavior of the magnetization at high fillings is affected by the parameters controlling the system anisotropy. We then outline the theoretical framework for studying the IEE through a self-consistent approach. Finally, we discuss expected outcomes in Rashba-like models.\\
\textit{DEE: Analytical and Numerical Calculation of Spin Density}- In this section, we present both analytical and numerical calculations of the DEE in the framework of the Rashba model, examining both isotropic and anisotropic cases. The DEE is a spin-to-charge conversion phenomenon, in which the application of an external electric field generates an electric current, leading to an accumulation of spin density perpendicular to, and co-planar with, the resulting current. The primary quantity of interest in this context is the spin susceptibility, which characterizes the efficiency of spin accumulation under the influence of the electric field. For the isotropic Rashba model, we analyze the behavior of spin susceptibility in systems where the Rashba spin--orbit interaction is uniform and directionally invariant. In the anisotropic Rashba model, we introduce not only a directional dependence in the SOC but also an anisotropy in the effective masses of the charge carriers. This allows to investigate how the combined effects of SOC anisotropy and effective mass anisotropy influence the spin susceptibility. \\
\textit{DEE in Isotropic Rashba Model}- Let us consider a Rashba system under the application of an external electric field $\mathbf{E}$. Within the Boltzmann framework, the expectation value of the magnetization $\mathbf{M}$, or the total spin density, at first order in the electric field can be obtained as
\begin{equation}
\mathbf{M}=- \mu_b \sum_{\mathbf{k}, \nu}|e|\left(\vec{\nu}^\nu({\mathbf{k}}) \cdot \mathbf{E}\right) \delta\left[\mathcal{E}^\nu(\mathbf{k})- \mathcal{E}_F \right]\langle\vec{\sigma}\rangle_{\mathbf{k}}^\nu,
\label{eqspin}
\end{equation}
where $\mathbf{k}$ is the quasi-momentum, $\nu=\pm$  is the index indicating the two chiral Fermi surfaces (see Figure~\ref{Edelstein}), and $\mu_b$ is the Bohr magneton. $\Vec{\nu}^{\nu}({\mathbf{k}})$= $\Bar{\tau}^{\nu}_{\mathbf{k}} \mathbf{v}^{\nu}({\mathbf{k}})$ indicates the mean free path, with the transport lifetime $\Bar{\tau}_{\mathbf{k}}^{\nu}$ and the group velocity $\mathbf{v}^{\nu}({\mathbf{k}})$= $\nabla_{\mathbf{k}} \varepsilon_{\mathbf{k}}^{\nu}$. ~\cite{johansson2016theoretical} 
The spin expectation value evaluated on the eigenstates is given by
\begin{equation}
    \langle \Vec{\sigma} \rangle_{\mathbf{k}}^{\pm} = \frac{1}{k} \Bigg ( 
    \begin{array}{c}
    \pm k_y \\
    \mp k_x \\
    0 \\
    \end{array}
    \Bigg ) = \Bigg ( 
    \begin{array}{c}
    \pm \sin(\theta)\\
    \mp \cos(\theta)\\
    0\\
    \end{array}
    \Bigg ),
    \end{equation}
where $\theta$ is the angle between the vector $\mathbf{k}$ and the $\hat{x}$ axis and $k=\sqrt{k_x^2+k_y^2}$. 
In a system as simple as the isotropic Rashba system, it is possible to carry out analytical calculations and derive the expression for the total spin density along $\hat{y}$.
Considering an initial electric field $\mathbf{E}= E_x \hat{x}$, when both chiral bands are occupied, i.e., in the High-Density Regime (HDR),  the spin density along the $\hat{y}$ direction is
\begin{equation}
\begin{split}
    m_y=& - \mu_b \sum_{\mathbf{k} \nu}  |e| \Bar{\tau}_{\mathbf{k}}^{\nu} ( \mathbf{v}^{\nu}({\mathbf{k}}) \cdot \mathbf{E}) \delta\left[\mathcal{E}^\nu(\mathbf{k})-\mathcal{E}_{\mathrm{F}}\right]\langle{\sigma_y}\rangle_{\mathbf{k}}^{\nu} = \\
   &\frac{\mu_b |e| E_x}{4 \pi} \big ( \Bar{\tau}_{+} k_F^+ - \Bar{\tau}_{-} k_F^- \big ), 
   \end{split}
   \label{spindensityHD}
\end{equation}
where $\Bar{\tau}_{+}$ and $\Bar{\tau}_{-}$ are the transport times, respectively, for the two energy bands, and 
\begin{equation}
\begin{cases}
      k_F^+= -k_0 + \sqrt{k_0^2 + 2mE_F}  \\
      k_F^-= +k_0 + \sqrt{k_0^2 + 2mE_F}
    \end{cases}
    \textrm{in HDR,}
\label{kHD}
\end{equation}
with $k_0=\alpha m$. For the complete derivation, refer to Appendix~\ref{calcolo_sigma_HD}.\\
Performing the same calculations in the Low-Density Regime (LDR), i.e., when only the lowest energy band is occupied, the spin density is described by the same expression as in Equation \eqref{spindensityHD} with  
\begin{equation}
    \begin{cases}
      k_F^+= + k_0 - \sqrt{k_0^2 + 2mE_F}  \\
      k_F^-= +k_0 + \sqrt{k_0^2 + 2mE_F}
    \end{cases}
    \textrm{in LDR.}
\label{kDSO}
\end{equation}
For the complete derivation, we refer the reader to Appendix~\ref{calcolo_sigma_LD}. So, the difference between the two regimes will lie in the explicit expressions of $\Bar{\tau}_{+}$, $\Bar{\tau}_{-}$, $k_F^+$, and $k_F^-$.\\
It is also possible to perform a numerical analysis. Defining the linear DEE as $m_j=\chi_{ij}E_i$, where $E_i$ is the electric field along the direction $\hat{i}$, then $\chi_{ij}$ is the Edelstein susceptibility. For simplicity, let us assume that $\Bar{\tau}_{+}=\Bar{\tau}_{-}=\tau$ is constant and $\mathbf{E}=E_x\hat{x}$. 
Explicitly, we can write 
\begin{equation}
    \chi_{xy}=-\chi_0 \sum_{\nu=\pm}\int d^2 \mathbf{k} \langle \sigma_y \rangle^{\nu}_{\mathbf{k}} \delta (\varepsilon_{\mathbf{k}}^{\nu}-\mu)v^{\nu}_x (\mathbf{k}),
    \label{eq:edel_suscept}
\end{equation}
where $\chi_0=\frac{\tau |e| \mu_b S_{\text{cell}}}{4\pi^2 a}$, with $a$ being the lattice parameter, $ S_{\text{cell}}$ the area of the unit cell, and $\tau=10^{-12}$~s the typical order of magnitude of the transport time in oxides ~\cite{trama2022tunable}.
As we see in Figure~\ref{rashba_iso_2} (right panel), the Rashba spin--orbit term alone is sufficient to induce an in-plane Edelstein response in the system.
\noindent Considering the expressions~\eqref{kHD} and~\eqref{kDSO} for the quasi-momentum at the Fermi energy, we can see how the behavior of the spin density in Figure~\ref{rashba_iso_2} (right panel) well represents the analytical results. 
Indeed, by substituting~\eqref{kHD} and~\eqref{kDSO} into~\eqref{spindensityHD}, in the HDR we obtain
\begin{equation}
  M_{y} = \frac{\mu_b|e| \tau} {2 \pi}  m \alpha  [\hat{z} \times \mathbf{E}]_y,
  \label{limite_spin_density_HD}
\end{equation}
so the spin density is constant and independent from $E_F$.\\
In the LDR, the expression for the spin density becomes 
\begin{equation}
  M_{y} = \frac{\mu_b|e| \tau}{2 \pi} \sqrt{(m^2 \alpha^2 + 2mE_F)} [\hat{z} \times \mathbf{E}]_y.
\end{equation}
For values of the Fermi energy around the band crossing, we can expand this expression for small values of $E_F$, 
\begin{equation}
  M_{y} = \frac{\mu_b|e| \tau}{2 \pi}  \biggl ( \alpha m + \frac{1}{2} \frac{E_F}{\alpha}\biggr ) [\hat{z} \times \mathbf{E}]_y.
\label{limite_spin_density_DSO}
\end{equation}
The spin density increases linearly with the Fermi energy. Note that for 
$\alpha\to 0$, the second term of the sum vanishes because the Fermi energy has a quadratic dependence on $\alpha$.\\
Returning to the numerical analysis of the problem, let us first consider Figure~\ref{rashba_isotropico 3}. 
The left panel of Figure ~\ref{rashba_isotropico 3} shows the Edelstein susceptibility as a function of the chemical potential for different values of $\alpha$. In particular, if $\alpha$ increases, also $\chi_{xy}/\chi_0$ increases as the value at which it reaches a plateau also increases. The right panel shows the susceptibility as a function of $\alpha$ at a fixed $\mu= 3.32 \cdot 10^{-2}$ eV; the susceptibility linearly increases with $\alpha$. This trend is also confirmed by the analytical expression discussed earlier (see Equation~\eqref{limite_spin_density_HD}), in which for high values of the chemical potential, it is linear with $\alpha$.
\textit{DEE in Anisotropic Case: The $C_{2v}$ Symmetry}- We now introduce an asymmetry into the model, characteristic, for example, of systems with  $C_{2v}$ symmetry. So, we can consider the anisotropy of both the effective mass $m$ and the Rashba parameter,  $r_m=\frac{m_x}{m_y}\ne 1$ 
and  $r_\alpha=\frac{\alpha_x}{\alpha_y}\ne 1$, respectively, where $x$ and $y$ correspond to the directions of the symmetry axes (see Figure ~\ref{pannelli_ani}). A mass anisotropy in the system can be easily achieved by uniaxial strain along a specific direction that can break the in-plane symmetry. On the other side,  the Rashba system is realized in a heterostructure or on a substrate, and the choice of substrate can modify the band structure asymmetrically. Moreover, the in-plane electric field can lead to asymmetric band warping, modifying the dispersion differently along the two axis. Finally, gate tuning can selectively alter the band curvature, especially in systems where the Rashba effect is controlled via external fields. 
The Hamiltonian in Equation~\eqref{hamiltonianaiso} becomes
\begin{equation}
    \hat{H} = \frac{\hbar^2 k_x^2}{2 m_x} + \frac{\hbar^2 k_y^2}{2 m_y} + \alpha_y k_y \hat{\sigma}_x - \alpha_x k_x \hat{\sigma}_y .
\end{equation}
In the following, we will first determine the Edelstein susceptibility numerically, and then derive the analytical expression as a function of anisotropies. By using the definition of the total spin susceptibility given in Equation~\eqref{eq:edel_suscept}, we can introduce anisotropy into the system by adjusting the parameters related to effective masses or spin--orbit coupling. In Figure~\ref{rashba_anitropico}, the Edelstein susceptibility is shown as a function of the chemical potential for different values of $r_m= m_y/m_x$ (left panel) and $r_{\alpha}= \alpha_y/\alpha_x$ (right panel).

\vspace{-6pt}
\begin{figure}[b!!]
    \begin{subfigure}[t]{0.5\textwidth}
    \centering
     \hspace{-6pt}   \includegraphics[width=\textwidth]{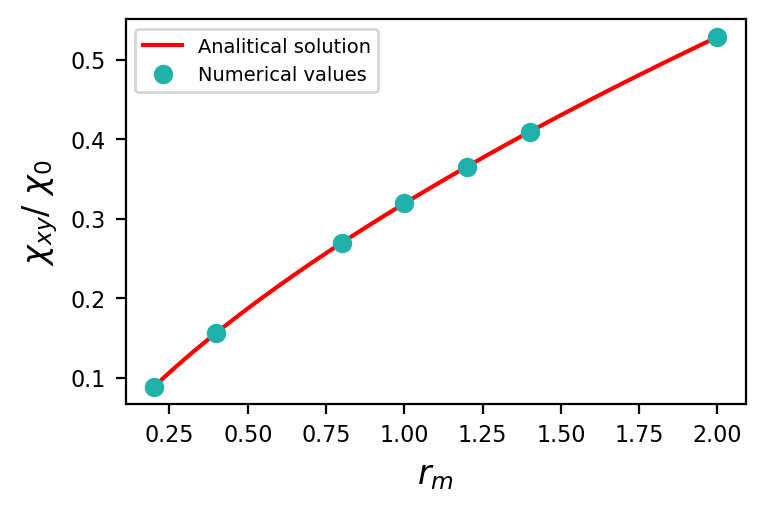}
        \label{4pannello-a}
    \end{subfigure}
    \begin{subfigure}[t]{0.5\textwidth}
    \centering
        \includegraphics[width=\textwidth]{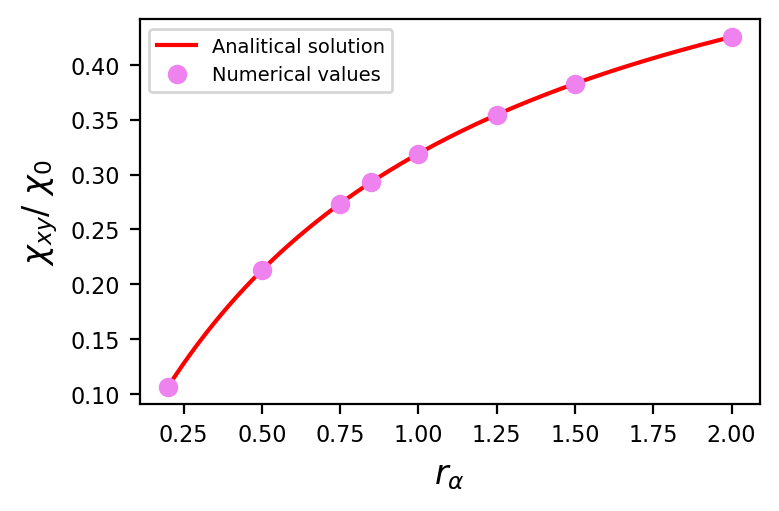}
        \label{4pannello-b}
    \end{subfigure}\vspace*{-8pt}
    \caption{Edelstein susceptibility $\chi_{xy}/\chi_0$ ( $\chi_0= 2.44 \times10^{2}$ $\mu_B $ ~\AA~V$^{-1}$) as a function of $r_m$ (\textbf{left panel}) and $r_{\alpha}$ (\textbf{right panel}), for a fixed chemical potential $\mu= 3.32 \cdot 10^{-2}$ eV. The points represent the numerical values obtained for the susceptibility, while the solid line represents the analytically derived expression.}
    \label{rashba_anisotro3}
\end{figure}

\hfill
\\
\noindent In the first case (left panel of Figure ~\ref{rashba_anitropico}), the SOC parameter $\alpha=\alpha_x=\alpha_y $ is fixed, while the ratio between the effective masses, $r_m$, changes. Specifically, $m_x= 0.152$ eV$^{-1}$ \AA$^{-2}$ is kept fixed, while $m_y$ is varied in such a way that $r_\alpha\in[0.2-2]$. 
In the right panel of Figure~\ref{rashba_anitropico}, the Edelstein susceptibility is shown as a function of the chemical potential for different values of $r_{\alpha}$. In this case, the effective masses are fixed, \mbox{$m_x=m_y=0.152$ eV$^{-1}$ \AA$^{-2}$},  \mbox{$\alpha_x=52$ eV \AA}, and the value of $\alpha_y$ is varied. Comparing Figures~\ref{rashba_iso_2} and~\ref{rashba_anitropico}, we observe that when $r_{m}$ and $r_{\alpha}$ are less than 1, the susceptibility is lower compared to the isotropic case. Conversely, when these ratios exceed 1, the susceptibility increases. This implies that one can boost the Edelstein effect by rendering $r_m$ and $r_\alpha$ greater than one. In Figure~\ref{rashba_anisotro3}, we show the Edelstein response as a function of $r_m$ and $r_{\alpha}$: we clearly see that the susceptibility increases with both these parameters. For small $r_m$ and $r_{\alpha}$ the Edelstein susceptibility is linearly increasing but tends to saturate for $r_{\alpha}$ much greater than one.  In particular, even in the anisotropic case, we are able to analytically extract the dependence of $\chi_{xy}/\chi_0$ on $r_m$ and $r_{\alpha}$ in the HDR:
\begin{equation}
\frac{\chi_{xy}}{\chi_0}(r_m) = \frac{4 \pi m_x \alpha r_m}{1 + \sqrt{r_m}}, \qquad
\frac{\chi_{xy}}{\chi_0}(r_\alpha) = \frac{4 \pi m    \alpha_x  r_\alpha }{1 + r_\alpha}.
\end{equation}
The full derivation of these expressions is given in Appendix~\ref{calcoli_analitici_anisotropo}.
In Figure~\ref{rashba_anisotro3}, points mark the numerical values of $\chi_{xy}/\chi_0$, while the solid red line corresponds to the analytical expressions provided below. The agreement between the numerical approach and the analytical estimate is quite evident.

\noindent Thus, the previous analysis of both isotropic and anisotropic Rashba systems highlights the interplay between the electric field, spin--orbit coupling, effective mass, and anisotropy parameters, providing a comprehensive understanding of spin dynamics in Rashba systems.

\begin{figure} [b!!]
\includegraphics[width= 8.5 cm]{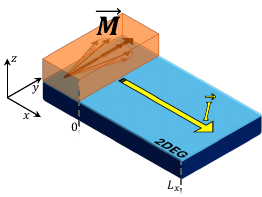}
\caption{Schematic view of the setup for observing the IEE. A 2DEG (in ligth-blue) is covered partly by a ferromagnetic metal (in orange). When a magnetization along $\hat{y}$ is near the electron gas, a spin bias along $\hat{y}$ occurs. Due to the IEE, an electric current is generated along $\hat{x}$ in the system.}  
\label{IEE_grafico_0}
\end{figure}

\newcommand{\kvec}{\mathbf{k}}
\textit{IEE: Inverse Edelstein Effect}- A key challenge in the field of spin-orbitronics remains the electrical detection of spin currents. One widely used approach is the spin Hall effect (SHE), where a pure spin current produces a detectable transverse charge current~\cite{hirsch1999spin, valenzuela2006direct, kimura2007room, werake2011observation, murakami2003dissipationless, jungwirth2012spin, kronmuller2007handbook, wunderlich2005experimental}. Despite its widespread application in spintronic experiments, the electrical signals generated by the SHE are typically small.
Another technique that has garnered increasing attention is the IEE~\cite{mahfouzi2014spin, shen2014microscopic}, where spin injection creates a non-equilibrium spin polarization, leading to a longitudinal charge current. The IEE has been observed in materials such as Bi ~\cite{isasa2016origin}, where it is attributed to RSOC at the interface. In contrast to the direct case, IEE does not involve the application of external fields. Instead, it can arise from the proximity of the system to a ferromagnet, where the magnetization of the ferromagnet, precessing, induces a spin accumulation in the system. This spin accumulation can be modeled as a spin injection along a specific direction. As a result, an electric current emerges in a direction perpendicular to that of the injected spin.
This phenomenon is fundamentally connected to Onsager reciprocal relations, which establish the symmetry between the DEE, where an electric field induces a spin polarization, and the IEE, where a spin accumulation generates an electric current~\cite{shen2014microscopic, gorini2012onsager}.\\
Figure~\ref{IEE_grafico_0} illustrates the setup for observing the IEE. A ferromagnet covers a part of an electron gas formed at the oxide interface characterized by a Rashba SOC. Then, a magnetic field is applied to pump a spin polarization in the lower gas, where a spin bias along $\hat{y}$ is generated. The system can be modeled as a Rashba electron gas, where the spin bias can be conveniently described by an operator ($-e\mu) \hat{\sigma} \cdot \hat{y}$~\cite{luo2016perfect}. 

To describe the IEE in 2D, we use a Boltzmann semiclassical approach. The inhomogeneous and stationary Boltzmann equation reads
\begin{equation}
     \textbf{v}_{\kvec} \cdot \frac{\partial f (\textbf{r},\textbf{k} )}{\partial \textbf{r}} = -\frac{f(\textbf{r},\textbf{k} ) - \left \langle f \right \rangle }{\tau},
\label{eq_bol_f}
\end{equation}
where the classical non-equilibrium distribution function $f(\textbf{r},\textbf{k} )$ of electrons in the system is a function of the phase-space point $(\textbf{r},\textbf{k})$, with $\textbf{k}$ and $\textbf{r}$ being the momentum and coordinate of an electron, respectively. In particular, since the distribution function is translational invariant along $\hat{y}$, the left term of the Boltzmann equation can be rewritten as $v_{x\kvec} \frac{\partial f(x,\kvec) }{\partial x}$. 
We can linearize the Boltzmann equation by writing 
\begin{equation}
   f_{x,\kvec}= f_0 + \bigg ( -\frac{\partial f_0}{\partial \varepsilon_{\kvec}} \bigg )g(x,\kvec), 
\label{f_distribution}
\end{equation}
where $f_0$ is the equilibrium Fermi distribution function and $g(x,\kvec)$ is a function that satisfies the same Boltzmann equation for the $f(x,\kvec)$. By substituting Equation~\eqref{f_distribution} into Equation~\eqref{eq_bol_f}, one obtains an equation for $g$: 
\begin{equation}
    v_{x \kvec}^{\nu}\frac{\partial g^{\nu}(x,\kvec)}{\partial x}= - \frac{g^{\nu}(x,\kvec)- \Bar{g}^{\nu}(x)}{\tau},
    \label{eqforg}
\end{equation}
where $\nu=\pm$ refers to the two chiral bands of the Rashba gas, and 
\begin{equation}
    \Bar{g}_k^{\nu}(x)= \frac{1}{2 \pi} \int\limits_0^{2\pi} d\theta \, g^{\nu}(x,  \kvec)
\end{equation}
is the angular average of the distribution function. The notation $g^{\nu}_k$ is chosen because it depends only on the magnitude $k$ of the quasi-momentum. We note that the external force term is absent in Equation ~\eqref{eq_bol_f}. Indeed, in the system we have considered, no external fields act at the macroscopic level, and the current—the quantity we aim to calculate—flowing along the sample from $x=0$ is due to the fact that for $x<0$, there is a magnetization caused by a ferromagnet. Therefore, when studying the distribution function from $x=0$, the Boltzmann equation should not include any external field terms. However, as we will see, the proximity to a ferromagnet will be taken into account through the boundary conditions imposed on the distribution function itself. One can easily recover this term by considering an alternative system in which, for example, an electric field $\textbf{E}$ is applied across the sample along $\hat{x}$, and by using the following transformation $ w(x,\kvec)= g(x,\kvec) - e V_L + e \int_0^x E(\xi) d\xi$ ~\cite{geng2016unified}. Even in this case, it can be demonstrated that the electric current density $I$ due to the IEE is invariant under this transformation and is given by  
\begin{equation}
\begin{split}
   I(x)=& \frac{2 e}{h} \int \textbf{v} w(x,\kvec) \bigg (-\frac{\partial f_0}{ \partial \varepsilon_{\kvec}} \bigg ) d^2\kvec \equiv \\ & \frac{2 e}{h} \int \textbf{v} g(x,\kvec) \bigg (-\frac{\partial f_0}{ \partial \varepsilon_k} \bigg ) d^2\kvec. 
\end{split}
\end{equation}
Thus, the functions $g(x,\kvec)$ and  $w(x, \kvec)$ play the same role but with different boundary conditions.\\
Assuming now that the distribution function at position $x=0$ still carries the same spin polarization as the reservoir, we can express $g^{\nu}(x, \kvec)$ as a function of the average spin density along $\hat{y}$. Using the eigenstates already found for the Rashba Hamiltonian, we find $\langle E_{\pm} | \sigma_y | E_{\pm} \rangle= \mp \cos(\theta_\kvec)$, where $\theta_{\kvec}$ is the angle formed by the quasi-momentum with the $k_x$-axis. The dependence of the function $g$ on $\kvec$ is implicit through the velocity $v^{\nu}_{x\kvec}$. For this reason, from now on, we will directly express the distribution function as $g(x, v^{\nu}_{x \kvec})$, and  the distribution function takes the form 
\begin{equation}
\begin{split}
    g(x=0,v_{x \kvec}^{\nu}>0) = & -eB  \langle E_{\pm} | \sigma_y | E_{\pm} \rangle= \\ & = \pm eB\cos(\theta_{\kvec}), \\
\end{split}
\end{equation}
where the index $\pm$ indicates, respectively, the two chiral energy bands, and $B$ is a constant that is proportional to the magnetization at $x=0$ and the dimensions of a potential.
Integrating the first-order linear differential Equation (\ref{eqforg}) and taking the boundary conditions for $g$ into consideration, it is possible to obtain a formal solution for the distribution function:
\begin{equation}
\begin{split}
    & g(x,v_{x\kvec}^{\nu})= \Theta(v_{x\kvec}^{\nu}) \Biggl \{ (\pm e B \cos{\theta_{\kvec}}) e^{-\frac{x}{v_{x\kvec}^{\nu} \tau}} + \\
    & \int\limits_0^x \bar{g}_k^{\nu}(\xi) e^{-\frac{x-\xi}{v_{x\kvec}^{\nu}\tau}} \frac{1}{v_{x \kvec}^{\nu} \tau} d\xi \Biggr\} + \\
      & \Theta(-v_{x \kvec}^{\nu}) \left\{ \int\limits_{L_x}^x \bar{g}_k^{\nu}(\xi) e^{-\frac{x-\xi}{v_{x \kvec}^{\nu}\tau}}  \frac{1}{v_{x \kvec}^{\nu} \tau}  d\xi \right\},
\end{split}
\label{autocons}
\end{equation}
where $\Theta$ is the unit step function. We note that the unknown function $\bar{g}_k^{\nu}(x)$ appears on the right-hand side, which needs to be solved first. By taking the local angular average on both sides of Equation~\eqref{autocons}, one can derive a self-consistent integral equation for $\bar{g}_k^{\nu}(x)$:
\begin{equation}
\begin{split}
    \bar{g}_k^{\nu}(x)= & \pm e B \biggl \langle \Theta(v_{x \kvec}^{\nu})   \cos{\theta_\kvec} e^{-\frac{x}{v_{x \kvec}^{\nu} \tau}} \biggl \rangle+ \\ & \biggl \langle \int\limits_0^x  \bar{g}_k^{\nu}(\xi) \Theta(v_{x \kvec}^{\nu}) e^{-\frac{x-\xi}{v_{x \kvec}^{\nu}\tau}} \frac{1}{v_{x \kvec}^{\nu} \tau} d\xi \biggl \rangle + \\ 
    & \biggl \langle   \int\limits_{L_x}^x \bar{g}_k^{\nu}(\xi) \Theta(-v_{x \kvec}^{\nu}) e^{-\frac{x-\xi}{v_{x \kvec}^{\nu}\tau}}  \frac{1}{v_{x \kvec}^{\nu} \tau}  d\xi \biggl \rangle,
    \end{split}
    \label{autocons2}
\end{equation}
which can be solved numerically. Equations~\eqref{autocons} and~\eqref{autocons2} constitute the exact solution of the present model. Once $\bar{g}_{\kvec}^{\nu}(x)$ is obtained, the distribution function can be calculated by using Equation~\eqref{autocons}. The electrical current is consequently given by
\begin{equation}
    I(x)= e L_y \sum\limits_{\nu} \int v_{x \kvec}^{\nu} g^{\nu}(x,v_{x \kvec}^{\nu}) \biggl ( -\frac{\partial f_0}{\partial \varepsilon_{\kvec}^{\nu}}   \biggl )k \, dk d \theta.
\label{corrente_IEE_formula}
\end{equation}
We point out that $\bar{g}_k^{\nu}(x)$ is essentially the relative change in the chemical potential in the non-equilibrium state, which is a slowly varying function of the position $x$. In Refs.~\cite{geng2017theory, geng2016unified}, it is demonstrated that a linear approximation $\bar{g}_k^{\nu}(x)= a^{\nu} + b^{\nu}x $, where the coefficients will depend on the $k$ magnitude, generally works very well. So, replacing $g(x, v_{x \kvec}^{\nu})$ with a linear expression for $\bar{g}_k^{\nu}(x)$, we can numerically solve the electric current in Equation~\eqref{corrente_IEE_formula} without a self-consistent approach. We report  the complete derivation to achieve the expressions for the coefficients $a$ and $b$ in Appendix~\ref{calcoli_IEE}. \\
We also include in Appendix ~\ref{continuità_corrente} a brief analysis of the current as a function of position, where we show that its continuity is satisfied, ensuring the conservation of the current in the system. We reserve further investigations of the IEE and the full self-consistent solution for future works.\\
\textit{Conclusions}- In this study, we have explored the phenomena associated with the Edelstein effect in the context of Rashba spin--orbit systems using a semiclassical Boltzmann approach. Through analytical and numerical approaches, we examined both the DEE and the IEE. For the DEE, we started from the Boltzmann equation in the stationary and homogeneous case, and we analyzed spin susceptibility in the isotropic Rashba model, revealing its dependence on the Rashba spin--orbit coupling parameter. We found that as the spin--orbit coupling parameter $\alpha$ increases, the spin susceptibility also increases, exhibiting a linear trend, a behavior that can be verified analytically. Introducing anisotropy in spin--orbit coupling and effective mass allowed us to elucidate the interplay of these factors, highlighting the distinct behaviors in systems with $C_{2v}$ symmetry. Even in the case of the anisotropic Rashba model, we are able to analytically find the explicit dependence of the spin susceptibility on two parameters: $r_m$, the ratio between the effective masses $m_y/m_x$, and $r_{\alpha}$, the ratio between the spin--orbit parameters $\alpha_y/\alpha_x$. In both cases, the spin susceptibility increases with the increase in the ratio. Specifically, this increase in $\chi_{xy}/\chi_0$ becomes faster as the ratio between the masses grows. For IEE, we developed a semiclassical Boltzmann framework to model the emergence of electric currents driven by spin accumulation, particularly in systems proximate to ferromagnetic materials. This approach, already used for systems such as topological insulators, was further developed by us for a 2D electron gas \cite{geng2016unified,geng2017theory}. By deriving a self-consistent solution to the Boltzmann transport equation, we studied how spin injection from ferromagnetic reservoirs generates electric currents perpendicular to the spin accumulation direction. {We aim to further investigate the inverse Edelstein effect in such systems, for example, by avoiding approximations where possible, obtaining analytical results, and testing whether the Onsager relations between the stationary homogeneous DEE and the stationary inhomogeneous IEE could provide insights for anisotropic systems.} We are setting up a method that is in the calibration phase.\\
\textit{Author Contributions}- All the authors discussed the methodology and validate the results; software and formal analysis, I.G. and M.T.; writing-original draft preparation, I.G. and M.T.; all the authors participated in the writing---review and editing; supervision, R.C., F.R. and C.G. All authors have read and agreed to the published version of the manuscript.\\
\textit{Funding}- The authors acknowledge support from  Horizon Europe EIC Pathfinder under the grant IQARO number
 101115190. R.C. and F.R. acknowledge funding from Ministero dell’Istruzione, dell’Università e della Ricerca
 (MIUR) for the PRIN project QUESTIONS (Grant No.PRIN P2022KWFBH) and STIMO (Grant No. PRIN 2022TWZ9NR). This work received funds from the PNRR MUR project PE0000023-NQSTI (TOPQIN and
 SPUNTO).\\
\textit{Data Availability Statement}- All the data are available under request to be sent to igaiardoni@unisa.it.\\
\textit{Acknowledgments}- The authors acknowledge Annika Johansson for the interesting discussion.\\
\textit{Conflicts of Interest:}- The authors declare no conflicts of interest.\\
\textit{Abbreviations}- The following abbreviations are used in this manuscript:\\
\noindent 
\begin{tabular}{@{}ll}
DEE & Direct Edelstein Effect\\
IEE & Inverse Edelstein Effect\\
HDR & High-Density Regime\\
LDR & Low-Density Regime
\end{tabular}

\clearpage

\appendix
\section[\appendixname~\thesection]{Analytical calculation in the isotropic case}
In this appendix, we explicitly present the calculation of the spin density along the $\hat{y}$ direction in the case of the isotropic Rashba model. In fact, the isotropy of the system allows an analytical calculation of the integral in Equation~\eqref{eqspin} in polar coordinates. The following calculations demonstrate the computation of spin susceptibility in two distinct cases. First, we consider the energy range defined as the High-Density Regime (HDR), in which both Rashba bands are involved in the transport. Next, we consider the case referred to as the Low-Density Regime (LDR), in which only the lower-energy band is involved.
\subsection[\appendixname~\thesection]{Analytical calculation of the spin susceptibility in HD}\label{calcolo_sigma_HD}
\noindent For the analytical calculation, we start from the definition of spin susceptibility,
\begin{equation}
\begin{split}
   M_{y} = - \mu_b \sum_{\mathbf{k} \nu} |e| \Bar{\tau}_{\mathbf{k}}^{\nu} ( \mathbf{v}_{\mathbf{k}}^{\nu} \cdot \mathbf{E}) \delta\left[\mathcal{E}^\nu(\mathbf{k})-\mathcal{E}_{\mathrm{F}}\right]\langle\vec{\sigma}\rangle_{\mathbf{k}}^{\nu}  
 \end{split}
\end{equation}
In the HDR, the Fermi energy is fixed above the band crossing, and the sum over $\nu= \pm 1$ corresponds to the sum over the two bands: the higher- and the lower-energy bands. From here, we move from the discrete to the continuous representation, so the summation over $\mathbf{k} $ is expressed as an integral over $\mathbf{k}$, which we directly write as an integral in polar coordinates $k, \theta$. We can also write explicitly the sum over $\nu$, changing the delta variable from energy to quasi-momentum, and substituting the spin expectation value $\langle \sigma_y\rangle^{\pm} = \mp \cos \theta$. So, the expression for the spin susceptibility takes this form
\begin{equation}
\begin{split}
     M_y =&-\mu_b |e| E_x \bigg [ \int \frac{dk d\theta}{(2 \pi)^2} k \Bar{\tau}_{k}^+ v_{x \mathbf{k}}^+ \frac{\delta( k - k_F^+)}{|v_{k_F}^+|} (- \cos \theta ) + \\ &\int \frac{dk d\theta}{(2 \pi)^2} k \Bar{\tau}_{k}^- v_{x \mathbf{k}}^-  \frac{\delta( k - k_F^-)}{|v_{k_F}^-|} (\cos \theta ) \bigg ],\\
    \end{split}
\end{equation}
where we indicate with $\Bar{\tau}_{k}$ the transport time that depends only on the magnitude of $\mathbf{k}$, with $v_{x \mathbf{k}}^\pm$ the component along $\hat{x}$ of the group velocity $v_{\mathbf{k}}^\pm$ and the superscripts $+$ and $-$ indicate which energy band one is referring to. Now, since $v_{x \mathbf{k}}^\pm = v_k^\pm \cos \theta$, the previous \mbox{expression becomes}
\begin{equation}
\begin{split}
    M_y=& -\mu_b |e| E_x \bigg [ \int\frac{dk d\theta}{(2 \pi)^2} k \Bar{\tau}_k^+ \frac{|v_k^+| \cos\theta}{|v_{k_F}^+|} \delta( k - k_F^+) (- \cos \theta ) + \\&\int \frac{dk d\theta}{(2 \pi)^2} k \Bar{\tau}_k^- \frac{|v_k^-| \cos\theta}{|v_{k_F}^-|}  \delta( k - k_F^-) (\cos \theta )\bigg ]=\\
    &-\frac{\mu_b |e| E_x}{2 \pi} \bigg [ \int dk k \Bar{\tau}_k^+ \delta( k - k_F^+) \int d\theta (-\cos^2 \theta) +\\ & \int dk k \Bar{\tau}_k^- \delta( k - k_F^-) \int d\theta (\cos^2 \theta) \bigg ].\\
 \end{split}
\end{equation}
By integrating over the quasi-momentum and angle, we obtain the expression for the spin density in the HDR:
\begin{equation}
\begin{split}
&M_y= \frac{\mu_b |e| E_x}{4 \pi} \Bar{\tau}_{+} k_F^+ - \frac{\mu_b |e| E_x}{4 \pi} \Bar{\tau}_{-} k_F^- ,
   \end{split}
\end{equation}
where $\bar{\tau}_{\pm}$ and $k_F^{\pm}$ here are only for the High-Density Regime.

\subsection[\appendixname~\thesubsection]{}{Analytical calculation of the spin susceptibility in LDR}
\label{calcolo_sigma_LD}
\noindent It is possible to perform the same calculation in the LDR, where the Fermi energy is fixed below the band crossing:
\vspace*{-6pt}
\begin{equation}
\begin{split}
   M_y = &- \mu_b \sum_{\mathbf{k} \nu} |e| \Bar{\tau}_{\mathbf{k}}^{\nu} ( \mathbf{v}_{\mathbf{k}}^{\nu} \cdot \mathbf{E}) \delta\left[\mathcal{E}^\nu(\mathbf{k})-\mathcal{E}_{\mathrm{F}}\right]\langle\vec{\sigma}\rangle_{\mathbf{k}}^{\nu} = \\
   &-\mu_b |e| E_x \bigg [ \int \frac{dk d\theta}{(2 \pi)^2} k \Bar{\tau}_k^+ v_{x \mathbf{k}}^+ \frac{\delta( k - k_F^+)}{|v_{k_F}^+|} (\cos \theta ) + \\ &\int \frac{dk d\theta}{(2 \pi)^2} k \Bar{\tau}_k^- v_{x \mathbf{k}}^-  \frac{\delta( k - k_F^-)}{|v_{k_F}^-|} (\cos \theta ) \bigg ]=\\
   & -\mu_b |e| E_x \bigg [ \int\frac{dk d\theta}{(2 \pi)^2} k \Bar{\tau}_k^+ \frac{(-|v_k^+| \cos\theta)}{|v_{k_F}^+|} \delta( k - k_F^+) ( \cos \theta ) + \\ &\int \frac{dk d\theta}{(2 \pi)^2} k \Bar{\tau}_k^- \frac{|v_k^-| \cos\theta}{|v_{k_F}^-|}  \delta( k - k_F^-) (\cos \theta )\bigg ]=\\
    &\frac{\mu_b |e| E_x}{2 \pi} \bigg [ \int dk k \Bar{\tau}_k^+ \delta( k - k_F^+) \int d\theta (\cos^2 \theta) - \\ 
    &\int dk k \Bar{\tau}_k^- \delta( k - k_F^-) \int d\theta (\cos^2 \theta) \bigg ]=\\
  &\frac{\mu_b|e| E_x}{4 \pi} \Bar{\tau}_{+} k_F^+ - \frac{\mu_b |e| E_x}{4 \pi} \Bar{\tau}_{-} k_F^-. 
   \end{split}
\end{equation}

The index $\nu$ now refers only to the lower-energy band, taking the value $+$ when considering the inner Fermi surface and $-$ when considering the outer one. $\bar{\tau}_{\pm}$ and $k_F^{\pm}$ here are the expression in the LDR.

\section[\appendixname~\thesection]{Analytical calculation in the anisotropic case}
\label{calcoli_analitici_anisotropo}
\noindent In this appendix, we present the analytical calculations for the spin susceptibility $\chi_{xy}/\chi_0$ generated by the DEE in a 2DEG. Specifically, even in the anisotropic case, in the HDR, we can analytically extract the dependence of $\chi_{xy}/\chi_0$ on the two parameters $r_m= m_y/m_x$ and $r_{\alpha}= \alpha_y/\alpha_x$.\\
We consider first the case where anisotropy in the system is introduced through the effective masses, looking at the derivation of $\chi_{xy}/\chi_0$ as a function of $r_m$. Starting from the expression for the spin susceptibility,
\begin{equation}
    \chi_{xy}=-\chi_0 \sum_{\nu=\pm}\int d^2\mathbf{k}\; \langle \sigma_y \rangle(\mathbf{k})\delta(\mathcal{E}_{\mathbf{k}}^\nu-\mu)v^\nu_x(\mathbf{k}),
    \label{eq:edel_suscept_2}
\end{equation}
it is possible to perform a change in variables from Cartesian to polar coordinates such that $k_x=k\cos{\theta}$ and $k_y=k\sin{\theta}$. Then, we can rewrite the components of the integral \eqref{eq:edel_suscept_2} in the new coordinates:
\begin{equation}
\begin{split}
    \mathcal{E}_{\mathbf{k}}^{\pm}= &
    \frac{k_x^2}{2m_x}+ \frac{k_y^2}{2m_y}+ \pm \alpha \sqrt{k_x^2+ k_y^2} = \\ &k^2 \bigg ( \frac{\cos^2{\theta}}{2m_x} + \frac{\sin^2{\theta}}{2m_x} \bigg) \pm \alpha k,
\label{energia_rm}
\end{split}
\end{equation}
\begin{equation}
    v^{\pm}_{x \mathbf{k}}= \frac{\partial \mathcal{E}_{\mathbf{k}}^{\pm}}{\partial k_x}= \frac{k_x}{m_x}\pm \frac{\alpha k_x}{\sqrt{k_x^2+ k_y^2}} = \frac{k \cos{\theta}}{m_x} \pm \alpha\cos{\theta} 
\label{vel_rm}
\end{equation}
\begin{equation}
    \langle \sigma_y \rangle^\pm= \mp \cos{\theta}
\label{sigma_rm}
\end{equation}
By substituting Equations ~\eqref{energia_rm}-~\eqref{sigma_rm} into   
 Equation~\eqref{eq:edel_suscept_2}, going from the delta in energy to the delta in the quasi-momentum $k$, and performing the products, the integral for the susceptibility is obtained in the following form:
\begin{equation}
\begin{split}
    &\chi_{xy}/\chi_0 =- \sum_{\nu=\pm}\int dkd\theta\; \frac{  k \langle \sigma_y^{\pm} \rangle v^\pm_{x \mathbf{k}} \delta(k- k_F^\pm)}{\bigg | k_F^\pm \bigg ( \frac{\cos^2{\theta}}{m_x} + \frac{\sin^2{\theta}}{m_x}\bigg ) \pm \alpha \bigg |}
    = \\
    &\int_0^{2\pi} -\frac{2 m_x m_y \alpha \cos^2\theta \left( m_y \left( -2 + \cos^2\theta\right) + m_x \sin^2\theta \right)}{\left( m_y \cos^2\theta + m_x \sin^2\theta \right)^2} d\theta,
\end{split}
\label{integrale_ani_rm}
\end{equation}
where $k_F^\pm $ are, respectively, the roots of the equation \\ $\mathcal{E}_{\mathbf{k}}^{\pm}- \mu=0$:
\begin{equation}
    \begin{split}
        k_F^\pm= \frac{\mp\alpha + \sqrt{\alpha^2 + 4 \mu \, t}}{2 \, t},
    \end{split}
\end{equation}
with $t= \frac{\cos^2\theta}{2 m_x} + \frac{\sin^2\theta}{2 m_y}$. 
The integral~\eqref{integrale_ani_rm} can be solved by invoking the residue theorem using the substitution $\cos(\theta)=\frac{z+z^{-1}}{2}$ and $\sin(\theta)=\frac{z-z^{-1}}{2i}$, where $z=e^{i\theta}$ and $d\theta=\frac{dz}{iz}$.
In this way, the integral~\eqref{integrale_ani_rm} reads $\frac{\chi_{xy}}{\chi_0}  = \int_{\gamma} I(z) dz$, with 
\begin{widetext}
\begin{equation}
    \mathcal{I}(z)=-\frac{2 i m_x m_y (1 + z^2)^2 \left(m_x (-1 + z^2)^2 - m_y (1 - 6 z^2 + z^4)\right) \alpha}{z (m_x-m_y)^2\left((z-z_1)(z-z_2)(z-z_3)(z-z_4)\right)^2},
    \label{eq:integrand_rm}
\end{equation}
\end{widetext}

where $\gamma$ is the anti-clockwise-oriented unitary circle, and $z_i$ are the second-order poles of the function, i.e., $z_1=-\sqrt{-1 + \frac{2 \sqrt{m_x}}{\sqrt{m_x} + \sqrt{m_y}}}$, $z_2=\sqrt{-1 + \frac{2 \sqrt{m_x}}{\sqrt{m_x} + \sqrt{m_y}}}$, $z_3=-\sqrt{-1 + \frac{2 \sqrt{m_x}}{\sqrt{m_x} - \sqrt{m_y}}}$, and $z_4=\sqrt{-1 + \frac{2 \sqrt{m_x}}{\sqrt{m_x} - \sqrt{m_y}}}$.
Since $z=0$, $z=z_1$, and $z=z_2$ are inside the unit circle, only they contribute to the integral. Therefore, we obtain
\begin{equation}
\begin{split}
    \frac{\chi_{xy}}{\chi_0} = &
    2i\pi \left( 
    \lim_{z \to 0} \left[ \mathcal{I}(z) \cdot z \right] + 
    \sum_{i=1}^2 \lim_{z \to z_i} \left[ \partial_z \left( \mathcal{I}(z) \cdot (z - z_i)^2 \right) \right]
    \right) =\\
    &\frac{4 \sqrt{m_x} m_y \pi \alpha}{\sqrt{m_x} + \sqrt{m_y}}.
\end{split}
\end{equation}
In terms of the ratio $r_m=m_y/m_x$, we obtain
\begin{equation}
    \frac{\chi_{xy}}{\chi_0}(r_m) = \frac{4 \pi \alpha  m_x  r_m}{1 + \sqrt{r_m}}.
\end{equation}
Considering anisotropy in the SOC parameter, i.e., $\alpha_x \neq \alpha_y$, we first recast the components of the integral \eqref{eq:edel_suscept_2} in the new coordinates,
\begin{equation}
\begin{split}
    \mathcal{E}_{\mathbf{k}}^{\pm}= &\frac{k_x^2 + k_y^2}{2m} \pm \sqrt{\alpha_x k_x^2 + \alpha_y k_y^2}=\\ & \frac{k^2}{2m} \pm k \sqrt{\alpha_x^2 \cos^2{\theta} + \alpha_y^2 \sin^2{\theta} }
\label{energia_ra}   
\end{split}
\end{equation}
\begin{equation}
\begin{split}
    v^{\pm}_{x \mathbf{k}} = &\frac{\partial \mathcal{E}_{\mathbf{k}}^{\pm}}{\partial k_x}= \frac{k_x}{m}\pm \frac{\alpha_x k_x}{\sqrt{\alpha_x^2 k_x^2+ \alpha_y^2 k_y^2}} = \\ & \frac{k \cos{\theta}}{m} \pm \frac{\alpha_x \cos{\theta} }{\sqrt{\alpha_x^2 \cos^2{\theta} + \alpha_y^2 \sin^2{\theta}}}
\label{vel_ra}
\end{split}
\end{equation}
\begin{equation}
    \langle \sigma_y \rangle = \mp \frac{\alpha_x \cos{\theta}}{\sqrt{\alpha_x^2 \cos^2{\theta} + \alpha_y^2 \sin^2{\theta}}} 
\label{sigma_ra}
\end{equation}
As in the previous case of anisotropy in the masses, we substitute Equations ~\eqref{energia_ra}-~\eqref{sigma_ra} into Equation~\eqref{eq:edel_suscept_2}: 
\begin{equation}
\begin{split}
    &\chi_{xy}/\chi_0 =- \sum_{\nu=\pm}\int dkd\theta\; \frac{ k \langle \sigma_y^{\pm} \rangle v^\pm_{x \mathbf{k}} \delta(k- k_F^\pm)}{\bigg | \frac{k_F^\pm}{m} \pm \sqrt{\alpha_x^2 \cos^2{\theta} + \alpha_y^2 \sin^2{\theta}} \bigg |}\\
    &\int_0^{2\pi} \frac{2 m \alpha_x \cos^2\theta \left( \alpha_y^2 + (\alpha_x - \alpha_y)(\alpha_x + \alpha_y) \cos(2 \theta) \right)}{\alpha_x^2 \cos^2\theta + \alpha_y^2 \sin^2\theta} d\theta,
\end{split}
\label{integrale_ani_ra}
\end{equation}
where $k_F^\pm$ are, respectively, the roots of the equation\\ $\mathcal{E}_{\mathbf{k}}^{\pm} - \mu=0$, i.e.,
\begin{equation}
    k_F^{\pm}= - m t^{'} + \sqrt{(t^{'2} m^2) + 2 \mu m},
\end{equation}
with $t^{'}= \sqrt{\alpha_x^2 \cos^2{\theta} + \alpha_y^2 \sin^2{\theta}}$.

\noindent The integral in the complex plane reads $\frac{\chi_{xy}}{\chi_0}=\int_{\gamma}\mathcal{I}(z)dz$, with
\begin{equation}
    \mathcal{I}(z)=-\frac{i (1 + z^2)^2 \left((1 + z^4) \alpha_x^2 - (-1 + z^2)^2 \alpha_y^2\right)}{2 z^3 (\alpha_x^2 - \alpha_y^2) (z - z_1) (z - z_2) (z - z_3)(z-z_4)},
    \label{eq:integrand_alpha}
\end{equation}
where the poles $z_i$ of $\mathcal{I}$ are $z_1 = -\frac{\sqrt{-\alpha_x - \alpha_y}}{\sqrt{\alpha_x - \alpha_y}}$, $z_2 = \frac{\sqrt{-\alpha_x - \alpha_y}}{\sqrt{\alpha_x - \alpha_y}}$, $z_3 = -\frac{\sqrt{-\alpha_x + \alpha_y}}{\sqrt{\alpha_x + \alpha_y}}$, and $z_4 = \frac{\sqrt{-\alpha_x + \alpha_y}}{\sqrt{\alpha_x + \alpha_y}}$.
Since only $z=0$, $z_3$, and $z_4$ lie within the unitary circle, the Edelstein susceptibility is given by
\begin{equation}
\begin{split}
    \frac{\chi_{xy}}{\chi_0}= &i\pi  
    \lim_{z \to 0} \partial_z^2 \left[ \mathcal{I}(z) \cdot z^3 \right] + \\ &
    2i\pi \sum_{i=3}^4 \lim_{z \to z_i} \left[\left( \mathcal{I}(z) \cdot (z - z_i) \right) \right]
    =\\&\frac{4 m \pi \alpha_x \alpha_y}{\alpha_x + \alpha_y},
\end{split}
\end{equation}
which can also be expressed in terms of $r_\alpha=\frac{\alpha_y}{\alpha_x}$ as
\begin{equation}
    \frac{\chi_{xy}}{\chi_0}=\frac{4  m  \pi  \alpha_x  r_\alpha }{1 + r_\alpha}.
\end{equation}

\section[\appendixname~\thesection]{Boltzmann Approach for the Inverse Edelstein Effect}
\label{calcoli_IEE}
\noindent We begin by substituting the linear approximation $\bar{g}_k^{\nu}(x)= a^{\nu} + b^{\nu}x $, where the coefficients depend on the magnitude of the quasi-momentum $k$, into Equation~\eqref{autocons2}:
\begin{equation}
\begin{split}
    a^{\nu} + b^{\nu}x = & \pm e B \biggl \langle \Theta(v_{x \kvec}^{\nu})   \cos{\theta_\kvec} e^{-\frac{x}{v_{x \kvec}^{\nu} \tau}} \biggl \rangle+ \\ &\biggl \langle \int\limits_0^x  (a + b \xi) \Theta(v_{x \kvec}^{\nu}) e^{-\frac{x-\xi}{v_{x \kvec}^{\nu}\tau}} \frac{1}{v_{x \kvec}^{\nu} \tau} d\xi \biggl \rangle + \\ 
    & \biggl \langle   \int\limits_{L_x}^x (a^{\nu} + b^{\nu}\xi) \Theta(-v_{x \kvec}^{\nu}) e^{-\frac{x-\xi}{v_{x \kvec}^{\nu}\tau}}  \frac{1}{v_{x \kvec}^{\nu} \tau}  d\xi \biggl \rangle
    \end{split}
    \label{autocons2_app}
\end{equation}
By solving the integrals in $d \xi$, we obtain the following equation for $a$ and $b$: 

\begin{equation}
\begin{split}
    \pm e B &\biggl \langle \Theta(v_{x \kvec}^{\nu})   \cos{\theta_\kvec} e^{-\frac{x}{v_{x \kvec}^{\nu} \tau}} \biggl \rangle + \\& \biggl \langle \Theta(v_{x \kvec}^{\nu}) e^{-\frac{x}{v_{x \kvec}^{\nu} \tau}} (-a^{\nu} + b^{\nu} \tau v_{x \kvec}^{\nu}) \biggl \rangle +\\ 
    &\biggl \langle \Theta(-v_{x \kvec}^{\nu})  e^{-\frac{x-L_x}{v_{x \kvec}^{\nu} \tau}} (-a^{\nu} + b^{\nu} \tau v_{x \kvec}^{\nu} - bL_x) \biggl \rangle =0
\end{split}
\label{Eqab}
\end{equation}

To determine the coefficients $a$ and $b$, one can choose two different values of $x$ in the above equation to obtain a couple of equations of $a$ and $b$. Noticing that the linear dependence of $\bar{g}(x)$ on $x$ is very well satisfied in the middle region of the sample, we choose $x=L_x/2$ and $x=L_x/2 + \Delta x$, and take the limit $\Delta x \rightarrow 0$. In the case of $x=L_x/2$, \mbox{Equation~\eqref{Eqab} becomes}
\begin{equation}
\begin{split}
    &- e B \biggl \langle \Theta(v_{x \kvec}^{\nu}) \cos{\theta_\kvec} e^{-\frac{L_x}{2 v_{x \kvec}^{\nu} \tau}} \biggl \rangle =  -a^{\nu} \biggl \langle \Theta(v_{x \kvec}^{\nu})  e^{-\frac{L_x}{2 v_{x \kvec}^{\nu} \tau}} \biggl \rangle + \\
    &b^{\nu}\tau v_F \biggl \langle \Theta(v_{x \kvec}^{\nu}) \cos{\theta_\kvec} e^{-\frac{L_x}{2 v_{x \kvec}^{\nu} \tau}} \biggl \rangle,
\end{split}
\end{equation}
where $v_F$ is the Fermi velocity, and we define
\begin{equation}
    \eta = \frac{\biggl \langle \Theta(v_{x \kvec}^{\nu}) \cos{\theta_\kvec} e^{-\frac{L_x}{2 v_{x \kvec}^{\nu} \tau}} \biggl \rangle}{\biggl \langle \Theta(v_{x \kvec}^{\nu}) e^{-\frac{L_x}{2 v_{x \kvec}^{\nu} \tau}} \biggl \rangle}.
\end{equation}
In the case of $x=L_x/2 + \Delta x$, where the velocity $v^{\nu}_{x \kvec}$ is expressed as $v_F^{\nu}\cos(\theta_\kvec)$, \linebreak Equation~\eqref{Eqab} can be written as
\begin{equation}
\begin{split}
    0= &-a^{\nu} \biggl \langle \Theta(-v_{x \kvec}^{\nu}) e^{\frac{L_x- 2\Delta x}{2 v_{x \kvec}^{\nu} \tau}}  \biggl \rangle + \\ &b^{\nu}\tau v_F^\nu \biggl \langle \Theta(v_{x \kvec}^{\nu}) \cos{\theta_\kvec}   e^{\frac{L_x- 2\Delta x}{2 v_{x \kvec}^{\nu} \tau}}  \biggl \rangle +\\
    &- b^{\nu} L_x \biggl \langle \Theta(-v_{x \kvec}^{\nu}) e^{\frac{L_x- 2\Delta x }{2 v_{x \kvec}^{\nu} \tau}}  \biggl \rangle,
\end{split}
\end{equation}
where we define
\begin{equation}
\begin{split}
    K =&  \frac{\biggl \langle \Theta(-v_{x \kvec}^{\nu}) \cos{\theta_\kvec} e^{\frac{L_x- 2\Delta x}{2 v_{x \kvec}^{\nu} \tau}} \biggl \rangle}{\biggl \langle \Theta(-v_{x \kvec}^{\nu}) e^{\frac{L_x- 2\Delta x }{2 v_{x \kvec}^{\nu} \tau}} \biggl \rangle} \xrightarrow{\Delta x \to 0} \\ &
    - \frac{\biggl \langle \Theta(v_{x \kvec}^{\nu}) \cos{\theta_\kvec} e^{-\frac{L_x}{2 v_{x \kvec}^{\nu} \tau}} \biggl \rangle} {\biggl \langle \Theta(v_{x \kvec}^{\nu}) e^{-\frac{L_x}{2 v_{x \kvec}^{\nu} \tau}} \biggl \rangle}
\end{split}
\end{equation}
Therefore, we end up with a system of equations for $a$ and $b$: 
\begin{equation}
\begin{split}
    a^{\nu}&= b^{\nu}(\tau v_F^{\nu} K - L_x)\\
    a^{\nu}&= \eta (eB) + b^{\nu} \tau v_F^{\nu} \eta 
\end{split}
\end{equation}
So, coefficients $a$ and $b$ are
\begin{equation}
\begin{split}
    a^{\nu}&= \frac{e B  (\tau v_F^{\nu} K - L_x)\eta}{\tau v_F^{\nu} (K-\eta) - L_x} = \frac{e B (L_x + \tau v_F \eta^2) }{ L_x+ 2 \tau v_F^{\nu} \eta } \\
    b^{\nu}&= \frac{e B \eta}{\tau v_F^{\nu}(K-\eta) - L_x} = -\frac{e B \eta}{L_x + 2 \tau v_F^{\nu} \eta}
\end{split}
\end{equation}

\section[\appendixname~\thesection]{Computation of the Inverse Edelstein Current}
\label{continuità_corrente}
\begin{figure}[b!!]
\includegraphics[width= 0.45\textwidth]{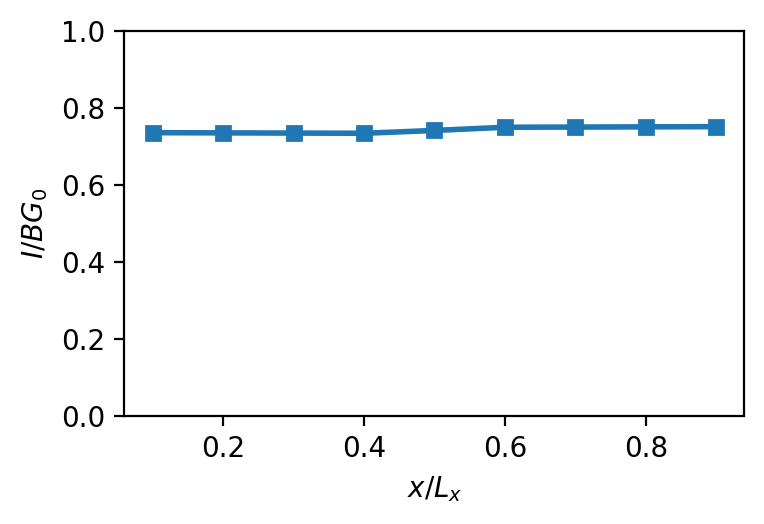}
\caption{Electric current, normalized by $B G_0$, as a function of $x/Lx$ at fixed chemical potential $0.17$ $eV$. In particular, $G_0= N_{ch}(e^2/h)$ with $N_{ch}= 4 k_F L_y/2 \pi$ as the number of conducting channels and $L_x \approx 10^{-4}$ as the length of the system along the $\hat{x}$ direction. The line is a visual guide.}
\label{IEE_grafico}
\end{figure}

\noindent In this section, we compute the IEE current through Equation~\eqref{corrente_IEE_formula} along $\hat{x}$, using a semi-analytical approach. We can use the linear approximation derived in Appendix~\ref{calcoli_IEE} and compute numerically the integrals in Equation~\eqref{autocons2}.
In Figure~\ref{IEE_grafico}, we show the behavior of the electric current normalized by $\mu G_0$, with $G_0$ as the Landauer--Büttiker conductance for a clean system~\cite{landauer1957spatial}, as a function of $x/L_x$ for a benchmark value of the chemical potential. 
\vfill  
\newpage  
\noindent The case represented here describes the current due to a magnetization of the system at $x=0$ in the diffusive regime, i.e., when the system length $L_x $ is much larger than the mean free path $l_f= v_F \tau$. It can be seen that the electric current remains constant along the $\hat{x}$ direction.

\bibstyle{unsrt}
\bibliography{Bibliography.bib}
\vspace{12pt}
\color{red}
\end{document}